\documentclass[twocolumn,aps,prl,showpacs]{revtex4}

\usepackage{epsfig}

\begin{document}

\title {Quantum Corrals, Eigenmodes and Quantum Mirages in s-wave Superconductors}
\author{Dirk K.~Morr and Nikolaos A. Stavropoulos}
\affiliation{Department of Physics, University of Illinois at
Chicago, Chicago, IL 60607}
\date{\today}
\begin{abstract}
We study the electronic structure of magnetic and non-magnetic
quantum corrals embedded in $s$-wave superconductors. We
demonstrate that a quantum mirage of an impurity bound state peak
can be projected from the occupied into the empty focus of a
non-magnetic quantum corral via the excitation of the corral's
eigenmodes. We observe an enhanced coupling between magnetic
impurities inside the corral, which can be varied through
oscillations in the corral's impurity potential. Finally, we
discuss the form of eigenmodes in magnetic quantum corrals.
\end{abstract}

\pacs{73.22.-f, 73.22.Gk, 72.10.Fk, 74.25.Jb}

\maketitle

The interaction of nanoscale impurity structures with fermionic
quantum many-body systems has led to the observation of a large
number of novel physical phenomena \cite{Man00,Hol01,Der02,Chi02}
over the last few years. In superconducting (SC) systems, quantum
interference of electronic waves that are scattered by a small
number of impurities can lead to the splitting of impurity states,
as observed in the one-dimensional chains of the high-temperature
superconductor YBa$_2$Cu$_3$O$_{6+x}$~\cite{Der02}, and discussed
theoretically for these chains~\cite{chains} as well as
two-dimensional (2D) $d_{x^2-y^2}$-wave \cite{dwave}
 and  $s$-wave superconductors (SSC) \cite{swave,Morr03b}. In more complex
impurity structures, such as quantum corrals, the existence of
discrete eigenmodes can be employed for the creation of {\it
quantum mirages}. This effect was beautifully demonstrated by
Manoharan {\it et al.}~\cite{Man00} who used the Kondo-resonance
of a magnetic impurity located in the focus of an elliptical
quantum corral as the ``electronic candle" whose quantum image was
projected into the empty focus. A nice theoretical explanation of
this phenomenon was subsequently provided in a series of articles
\cite{theory}.

Of particular interest is the possibility that nanoscale impurity
structures can provide insight into the nature of complex
electronic systems. In general, one expects that strong electronic
correlations or changes in the electronic structure due to broken
symmetries affect the spatial patterns of eigenmodes in quantum
corrals and provide novel ``electronic candles" whose
spectroscopic signatures can be projected. As a first step in the
investigation of this idea, we study in this Letter quantum
corrals that are embedded in an $s$-wave superconductor with
non-trivial correlations arising from particle-hole mixing. We
consider a variety of quantum corrals consisting of non-magnetic
or magnetic impurities with constant or oscillating scattering
potentials. Magnetic impurities that are placed inside the corral
induce bound states whose spectroscopic signature are peaks in the
density-of-states (DOS). We utilize these peaks as the
``electronic candle" to investigate the corral's electronic
properties. We demonstrate that by placing a magnetic impurity in
one of the corral's foci, a quantum image of its bound state peaks
is projected into the empty focus via the excitation of the
corral's eigenmodes. These eigenmodes also lead to an enhanced
coupling between magnetic impurities inside the corral. We
illustrate how the spatial pattern of eigenmodes can be changed
through oscillations in the corral's impurity potential or the
relative alignment of impurity spins in magnetic corrals. These
results provide a new tool for manipulating the interaction
between magnetic impurities, a topic of great current interest in
the field of spin electronics and quantum information
technology~\cite{spinQC}.

To study the electronic structure of quantum corrals we employ a
$\hat{T}$-matrix formalism \cite{chains,Morr03b,Shiba68,Balatsky}
which was generalized to describe electronic scattering off a
large number of impurities. In a fully gapped SSC, magnetic
impurities with spin $S$ can be treated as classical variables
\cite{Shiba68} since no Kondo-effect exists for sufficiently small
coupling between the impurity and delocalized electrons
\cite{Sat92}, in full agreement with experiment \cite{Yaz97}. In
the Nambu-formalism, the electronic Greens function in the
presence of $N$ impurities located at ${\bf r}_i$ is
\begin{eqnarray}
\hat{G}({\bf r},{\bf r'},\omega_n)&=&\hat{G}_0({\bf r},{\bf r'},\omega_n) \nonumber \\
& & \hspace{-2cm} +\sum_{i,j=1}^N \hat{G}_0({\bf r},{\bf
r}_i,\omega_n)\hat{T}({\bf r}_i,{\bf r}_j,\omega_n)\hat{G}_0({\bf
r}_j,{\bf r'},\omega_n) \ , \label{Ghat}
\end{eqnarray}
where one has for the $\hat{T}$-matrix
\begin{eqnarray}
\hat{T}({\bf r}_i,{\bf r}_j,\omega_n)&=&\hat{V}_i \, \delta_{i,j}
  \nonumber \\
& & \hspace{-1.5cm} +\hat{V}_i \, \sum_{l=1}^N \hat{G}_0({\bf
r}_i,{\bf r}_l,\omega_n)\hat{T}({\bf r}_l,{\bf r}_j,\omega_n) \ ,
\label{T1}
\end{eqnarray}
and
\begin{eqnarray}
& & \hat{G}^{-1}_0({\bf k},i\omega_n)=\left[ i\omega_n \tau_0 -
\epsilon_{\bf k} \tau_3 \right] \sigma_0 + \Delta_0 \tau_2
\sigma_2 ; \nonumber \\
& & \hat{V}_i=\frac{1}{2} \left(U_i \sigma_0 + J_iS \sigma_3
\right)\tau_3 \ . \label{VG}
\end{eqnarray}
$\hat{G}_0({\bf k},i\omega_n)$ is the electronic Greens function
of the unperturbed (clean) system in momentum space, $ {\bf
\sigma}$, ${\bf \tau}$ are the Pauli-matrices in spin and
Nambu-space, respectively, and $\Delta_0$ is the SC gap. We assume
that the corral is located at the surface of an SSC, and thus
consider for simplicity a 2D electronic system. Its normal state
dispersion is given by $\epsilon_{\bf k}= k^2/2m-\mu$ ($\hbar=1$),
where $\mu=k_F^2/2m$ is the chemical potential and $k_F=\pi/2$ is
the Fermi wave-vector (lattice constant $a_0=1$). We set
$1/(m\Delta_0)=30$, yielding a SC coherence length of
$\xi_c=k_F/(m\Delta_0)=15\pi$. Moreover, $\hat{V}_i$ is the
scattering matrix at ${\bf r}_i$, with $U_i$ and $J_i$ being the
potential and magnetic scattering strengths of the impurity.
Unless otherwise noted, we take for definiteness
$U_i/2\Delta_0=30$ ($J_i=0$) for non-magnetic impurities, and
$J_iS/2\Delta_0=30$ ($U_i=0$) for magnetic impurities. These
values are taken to demonstrate the qualitative features of our
results which are robust against changes in the scattering
strengths or in the form of the fermionic dispersion. The DOS,
$N({\bf r},\omega)$, is obtained from a numerical computation of
Eqs.(\ref{Ghat}) and (\ref{T1}) with $N({\bf
r},\omega)=A_{11}+A_{22}$, $A_{ii}({\bf r},\omega)=-{\rm Im}\,
\hat{G}_{ii}({\bf r},\omega+i\delta)/ \pi$, and $\delta=0.02
\Delta_0$.

%
% Figure 1
%
\begin{figure}[t]
\epsfig{file=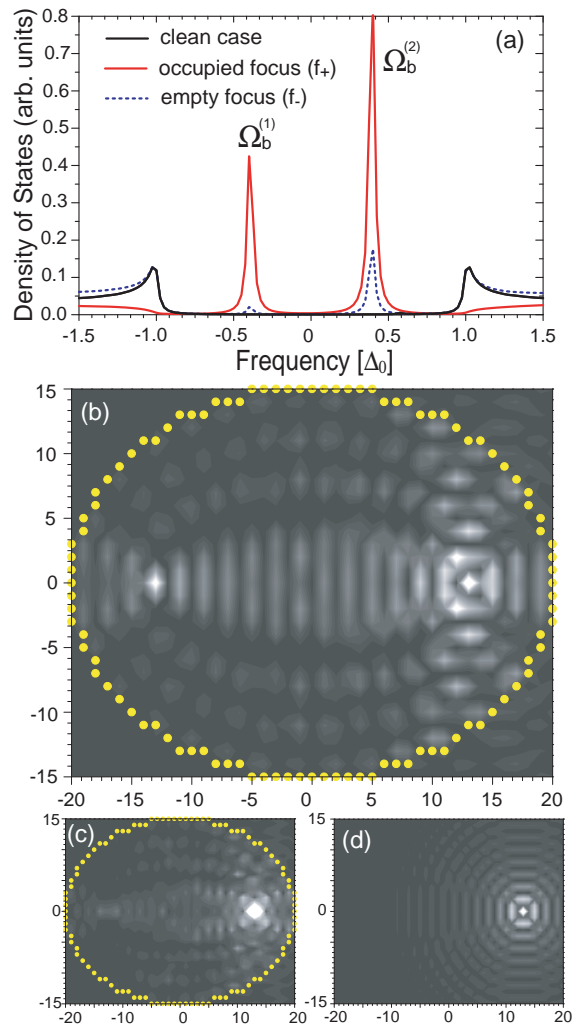,width=7.5cm} \caption{{\it (a)} DOS in the
occupied (red line) and empty focus (dotted blue line) for a
non-magnetic corral with a magnetic impurity located at $f_+$.
Black line: DOS of the clean system. {\it (b,c)} DOS inside the
quantum corral (the filled yellow circles represent the corral
impurities) for {\it (b)} $\Omega^{(2)}_b/\Delta_0= 0.4$, {\it
(c)} $\Omega^{(1)}_b/\Delta_0=-0.4$.  {\it (d)} DOS at
$\Omega^{(2)}_b/\Delta_0= 0.575$ for a magnetic impurity at $f_+$
without corral.} \label{corral2015}
\end{figure}
We first study an elliptical corral with semi-axes $a=20$, $b=15$,
and eccentricity $e=\sqrt{7}/4$, that consists of  100
non-magnetic impurities. Defining the center of the corral as
$(0,0)$, we place a magnetic impurity in the corral's focus at
$f_{+}=(13, 0)$, while leaving the other focus at $f_{-}=(-13,0)$
empty. In Fig.~\ref{corral2015}a, we present the DOS in the foci
at $f_{\pm}$. As expected, the magnetic impurity induces a bound
state resulting in a particle- and hole-like peak in the DOS at
frequencies $\Omega_b^{(1,2)}/\Delta_0=\mp 0.4 $, respectively. A
quantum mirage of these peaks emerges in the empty focus at $f_-$.
The formation of this quantum image through excitations of the
corral's eigenmodes becomes evident when we consider the spatial
DOS pattern at $\Omega_b^{(2)}$ as shown in
Fig.~\ref{corral2015}b. All plots of the spatial DOS shown in the
following possess the same intensity scale to facilitate a direct
comparison, with light (dark) color indicating a large (small)
DOS. In addition to the quantum mirage, we observe DOS
oscillations, representing the excited eigenmodes of the corral.
By increasing the corral's impurity potential \cite{Morr03c}, the
eigenmodes as well as the impurity bound state become more
confined inside the corral. The presence of two bound state peaks
at $\Omega_b^{(1,2)}$ that arise from particle-hole mixing in the
SC state allows us to study eigenmodes at different excitation
energies. Note, e.g., that the spectral weight in the DOS at
$\Omega_b^{(1)}$ (Fig.~\ref{corral2015}c) is much more
concentrated around $f_+$ than at $\Omega_b^{(2)}$
(Fig.~\ref{corral2015}b) concomitant with a weaker excited
eigenmode and quantum mirage.  It was noted earlier \cite{theory},
that eigenmodes can only be excited if the excitation (via the
impurity bound state) takes place at a position where the spectral
weight of the eigenmode is large, and if the excitation energy,
i.e., $\Omega_b^{(1,2)}$, is close to the eigenmode's energy. For
$U_i=\infty$, no eigenmodes exist inside the SC gap, and the
eigenmodes closest to the Fermi energy and large spectral weight
close to the foci are located at $\Omega^{(\pm)}_m/\Delta_0=\pm
1.1$. Since the mode's amplitude at $\Omega^{(+)}_m$ is
considerably larger than that at $\Omega^{(-)}_m$, the DOS
oscillations and thus the quantum mirage at $\Omega_b^{(1)}$ are
weaker, leading to a concentration of spectral weight around
$f_+$. This demonstrates that the eigenmodes act as ``waveguides"
for the projection of the bound state peaks into a quantum image
\cite{Man00,theory}.

The effect of a corral on the DOS strongly depends on the ratio of
the decay length, $\xi_d=\xi_c/\sqrt{1-(\Omega_b/\Delta_0)^2}$,
and the corral's semi-axes. In the above case,
$\Omega^{(1,2)}_b/\Delta_0=\pm 0.4$, $\xi_d=16.4 \pi \gg a,b$ and
the spatial DOS pattern at $\Omega^{(2)}_b$ is significantly
different for a magnetic impurity at $f_+$ with
(Fig.~\ref{corral2015}b) and without a corral
(Fig.~\ref{corral2015}d). Note that the presence of a corral also
shifts $\Omega^{(1,2)}_b$ from $\Omega^{(1,2)}_b/\Delta_0=\pm
0.575$ (no corral) to $\Omega^{(1,2)}_b/\Delta_0=\pm 0.4$. In
contrast, if $\xi_d \ll a,b$ the bound state wave-function at the
position of the corral wall is exponentially suppressed, no
eigenmodes can be excited, and the spatial DOS pattern of a single
magnetic impurity remains unchanged in the presence of a quantum
corral \cite{Morr03c}. Note, however, that $\xi_d$ can become
arbitrarily large by increasing $JS$ such that
$|\Omega^{(1,2)}_b|/\Delta_0 \rightarrow 1$.

%
% Figure 2
%
\begin{figure}[t]
\epsfig{file=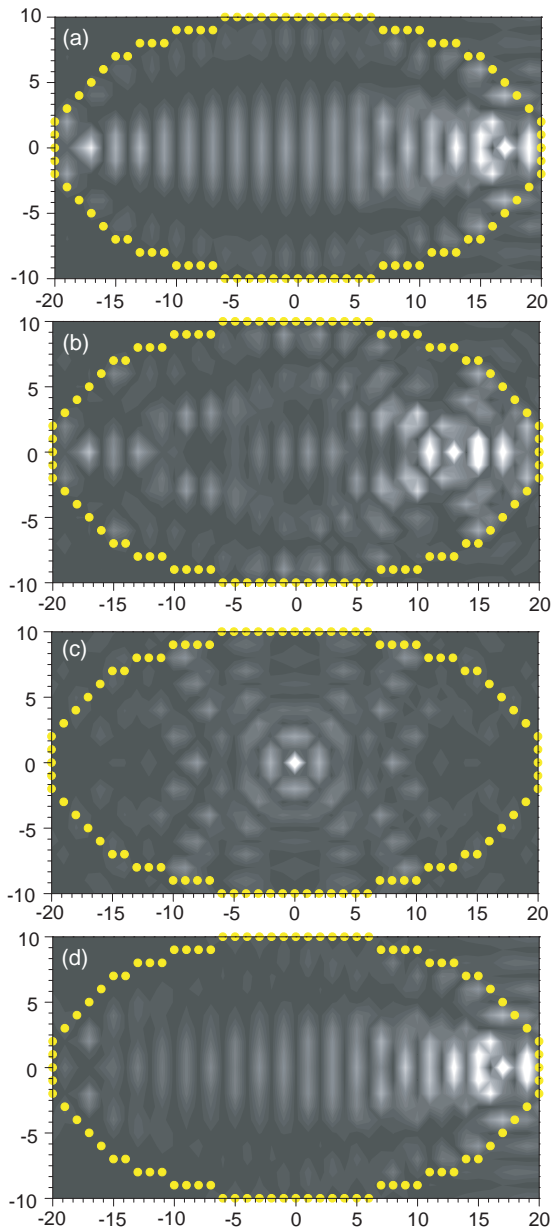,width=7.5cm} \caption{DOS at
$\Omega^{(2)}_b$ for a magnetic impurity located at {\it (a)}
$f_+=(17,0)$, {\it (b)} ${\bf r}=(13,0)$, and {\it (c)} ${\bf
r}=(0,0)$. {\it (d)} DOS for an oscillating impurity potential
along the corral with $U(\phi)=U_0\cos\phi$ and
$U_0/2\Delta_0=30$.} \label{corral2010}
\end{figure}
We next study the evolution of the DOS pattern when the magnetic
impurity is moved off the focus for a corral with 88 impurities,
$a=20$, $b=10$, eccentricity $e=\sqrt{3}/2$ and $f_{\pm}=(\pm
17,0)$. In Fig.~\ref{corral2010}a we present the DOS at
$\Omega^{(2)}_b/\Delta_0= 0.475$ for a magnetic impurity located
at $f_{+}$. The DOS exhibits a similar pattern, including a
quantum mirage at $f_{-}$, as in Fig.~\ref{corral2015}b, albeit
with only one instead of three ``side wings" in the excited
eigenmode. When we move the impurity off the focus to $(13,0)$
(Fig.~\ref{corral2010}b), the bound state energy increases to
$\Omega^{(1,2)}_b/\Delta_0=\mp 0.5$, and only a much weaker
quantum mirage emerges at $f_-$. The DOS pattern changes
significantly when the magnetic impurity is located at the center
of the corral at $(0,0)$ (Fig.~\ref{corral2010}c) with
$\Omega^{(1,2)}_b/\Delta_0= \mp 0.725$. Thus, changing the
location of the excitation, i.e., the magnetic impurity, leads to
different excited eigenmodes and a simultaneous shift in
$\Omega^{(1,2)}_b$. In Fig.~\ref{corral2010}d we plot the DOS in
the presence of an oscillating impurity potential along the
corral's wall with $U(\phi)=U_0\cos\phi$, $U_0/2\Delta_0=30$, and
$\phi$ being the angle between the x-axis and the line connecting
the center of the ellipse with the impurity. This oscillating
potential, which might arise, e.g., from charge oscillations in
the corral's wall, weakens the eigenmodes, particularly along the
vertical axis of the corral where $U(\phi)$ is small, and almost
completely destroys the quantum mirage of the bound state peak.

%
% Figure 3
%
\begin{figure}[t]
\epsfig{file=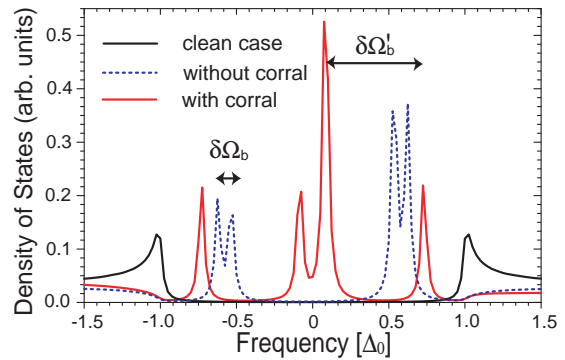,width=7.5cm} \caption{Splitting of the bound
state peaks in the DOS at $f_-=(-13,0)$ with and without a quantum
corral for two magnetic impurities with parallel spins.}
\label{2imp}
\end{figure}
It was argued in Ref.~\cite{swave,Morr03b} that quantum
interference effects between two magnetic impurities leads to the
formation of bonding and antibonding bound states and thus to a
frequency splitting of the bound state peaks. We find that this
splitting can be enhanced if the magnetic impurities are placed in
the foci of a corral. We consider the same corral as in
Fig.~\ref{corral2015} and assume for definiteness that the spins
of the impurities are parallel, however, qualitatively similar
results are obtained for arbitrary angle between the impurity
spins. As shown in Fig.~\ref{2imp}, in the absence of a quantum
corral, the energy splitting of the bound state peaks is small,
$\delta \Omega_b/\Delta_0 = 0.1$, due to the large distance
between the impurities. However, in the presence of the corral the
splitting increases to $\delta \Omega'_b/\Delta_0=0.65$, implying
that the coupling between the two magnetic impurities is enhanced
(in the normal state, this effect was discussed in
Ref.~\cite{interaction}). Note that for an oscillating impurity
potential, as that discussed in Fig.~\ref{corral2010}d, the
splitting decreases to $\delta \Omega''_b/\Delta_0 = 0.175$.

\begin{figure}[t]
\epsfig{file=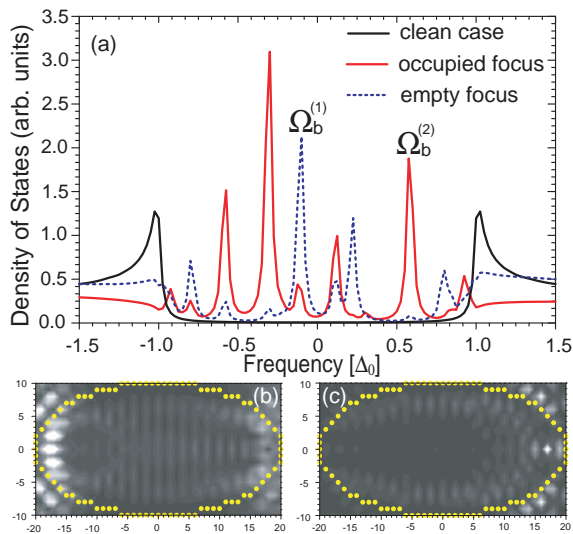,width=7.5cm} \caption{{\it (a)} DOS at
$f_{\pm}$ for a magnetic corral. Spatial DOS pattern for {\it (b)}
$\Omega^{(1)}_b/\Delta_0=-0.1$ and {\it (c)}
$\Omega^{(2)}_b/\Delta_0=0.575$ (see {\it (a)}).} \label{MC1}
\end{figure}

Finally, we present in Fig.~\ref{MC1} the DOS inside a quantum
corral ($a=20$, $b=10$) consisting of 88 magnetic impurities. The
corral's impurity spins are antiferromagnetically aligned and an
additional magnetic impurity is located at $f_+=(17,0)$.
%
% Figure 4
%
Since the bound states associated with the magnetic impurities are
coupled, the DOS exhibits a large number of peaks
(Fig.~\ref{MC1}a), and we do not observe a well defined quantum
image in the empty focus. The DOS pattern changes qualitatively
between different bound state energies, as shown in
Fig.~\ref{MC1}b and c where we present the DOS at
$\Omega^{(1)}_b/\Delta_0=-0.1$ and $\Omega^{(2)}_b/\Delta_0=0.575$
(see Fig.~\ref{MC1}a). For $\Omega^{(1)}_b$ the spectral weight of
the excited eigenmode is predominantly centered around the empty
focus, while for $\Omega^{(2)}_b$ most of the spectral weight is
located in the vicinity of the occupied focus. Thus, in contrast
to non-magnetic corrals, the eigenmodes in a magnetic corral can
be concentrated near the occupied {\it or} empty focus. For
ferromagnetic alignment of the corral spins (not shown) there also
exists a large number of bound state peaks in the DOS
\cite{Morr03c}. However, the spatial distribution of spectral
weight associated with these peaks is much more homogeneous,
likely due to the stronger scattering for ferromagnetically
aligned spins \cite{Morr03b} and the resulting larger number of
excited eigenmodes.

Finally, we comment on the relevance of pair-breaking effects near
magnetic impurities. Since the suppression of the SC gap close to
a single magnetic impurities does not alter the DOS's qualitative
features \cite{Balatsky}, our results presented in
Figs.~\ref{corral2015}-\ref{2imp} are likely robust against
pair-breaking effects. In magnetic corrals, on the other hand, the
suppression of the SC gap is potentially relevant and its
magnitude could depend on the ferro- or antiferromagnetic
alignment of the corral's spins. Work is currently under way to
study the significance of pair-breaking effects and gap
suppression in both cases \cite{Morr03c}.

In summary, we demonstrate that a quantum mirage of an impurity
bound state peak can be projected from the occupied into the empty
focus of a non-magnetic quantum corral via the excitation of the
corral's eigenmodes. We observe an enhanced coupling between
magnetic impurities inside the corral, which can be varied through
oscillations in the corral's impurity potential. This provides a
novel tool to manipulate the interaction between magnetic
impurities, a topic of great relevance in spin electronics and
quantum information technology \cite{spinQC}.

We would like to thank J.C. Davis and A. de Lozanne for
stimulating discussions.

\end{document}